\documentclass{article}

\usepackage{arxiv}

\usepackage[utf8]{inputenc} 
\usepackage[T1]{fontenc}    
\usepackage{hyperref}       

\usepackage{url}            
\usepackage{booktabs}       
\usepackage{amsfonts}       
\usepackage{nicefrac}       
\usepackage{microtype}      
\usepackage{lipsum}		
\usepackage[pdftex]{graphicx}
\usepackage{doi}
\usepackage[cmex10]{amsmath}
\usepackage{algorithmic}
\usepackage{array}
\usepackage[caption=false,font=footnotesize]{subfig}
\renewcommand{\caption}[2][\relax]{\MYoriglatexcaption[#2]{#2}}
\usepackage{subfloat}
\usepackage{dcolumn}
\usepackage{bm}
\usepackage{float}
\floatstyle{plaintop}
\restylefloat{table}
\usepackage{tabularx}     
\usepackage{siunitx}[=2021-04-09]
\usepackage{courier}
\usepackage{url}

\usepackage{cite}
\usepackage{etoolbox}
\usepackage[dvipsnames]{xcolor}
\usepackage{xspace}
\usepackage[normalem]{ulem}
\usepackage{cancel}

\DeclareRobustCommand{\ed}[2][blue]{\textcolor{#1}{#2}}

\title{Modeling dispersive silver in the electrodynamic lattice-Boltzmann method using complex-conjugate pole-residue pairs}

\date{September 7, 2022}	

\author{ \href{https://orcid.org/0000-0003-3493-527X}{\includegraphics[scale=0.06]{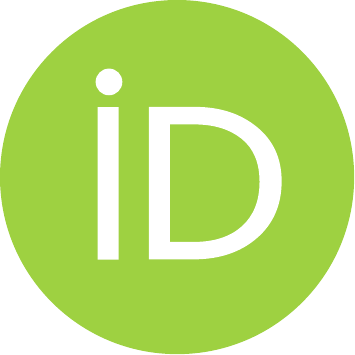}\hspace{1mm}Cael~Warner}\thanks{} \\
	Department of Electrical Engineering\\
	The University of British Columbia\\
	Kelowna, BC V1V1V7 \\\href{mailto:cael.warner@ubc.ca}{cael.warner@ubc.ca} \\
	\And
	\href{https://orcid.org/0000-0000-0000-0000}{\includegraphics[scale=0.06]{orcid.pdf}\hspace{1mm}Lo\"{i}c~Markley} \\
	Department of Electrical Engineering\\
	The University of British Columbia\\
	Kelowna, BC V1V1V7 \\
 \href{mailto:loic.markley@ubc.ca}{loic.markley@ubc.ca}. \\
 \And
	\href{https://orcid.org/0000-0000-0000-0000}{\includegraphics[scale=0.06]{orcid.pdf}\hspace{1mm}Kenneth~J.~Chau} \\
	Department of Electrical Engineering\\
	The University of British Columbia\\
	Kelowna, BC V1V1V7 \\
  \href{mailto:Kenneth.Chau@ubc.ca}{Kenneth.Chau@ubc.ca}
}

\renewcommand{\shorttitle}{Lattice-Boltzmann method for modeling one-dimensional dispersive silver}

\hypersetup{
pdftitle={Modeling dispersive silver in the electrodynamic lattice-Boltzmann method using complex-conjugate pole-residue pairs},
pdfsubject={physics.comp-ph},
pdfauthor={Cael~Warner, Lo\"{i}c~Markley, Kenneth~J.~Chau},
pdfkeywords={Dispersion, electromagnetism, lattice-Boltzmann, broadband, silver slab, time domain},
}

\begin{document}
\maketitle

\begin{abstract}
 The polarization density of a broadband electrodynamic lattice-Boltzmann method (ELBM) is generalized to represent frequency-dispersion of materials interacting with electromagnetic waves. The frequency-dependent refractive index and extinction coefficient are modeled using complex-conjugate pole-residue pairs in an auxiliary-differential-equation (ADE). Electric and magnetic fields are evaluated on a single lattice, ensuring a stable numerical solution up to the Nyquist limit. The electric and magnetic fields from the ELBM are compared with the electric and magnetic fields from the finite-difference-time-domain (FDTD) method.  Accurate transmittance of a 100 \si{nm} silver slab is extracted from the transmitted power spectrum of a broadband Dirac-delta wave-function for photon energies ranging from 0.125-5 \si{eV}. Given this capability, the ELBM with an ADE is an accurate and computationally efficient method for modeling broadband frequency-dispersion of materials.
\end{abstract}


\section{Introduction}
The electrodynamic lattice-Boltzmann method (ELBM) has been implemented for modeling interactions of electromagnetic waves with uniform dielectric \cite{Mendoza2010,Hanasoge2011,Hauser2017,Hauser2019,hauser2022simulation}, conductive \cite{Mendoza2010,Hauser2017}, non-linear and dispersive media \cite{Mendoza2011,Chen2013}. These methods can have superior accuracy and/or computational efficiency compared with the finite-difference-time-domain (FDTD) method \cite{KaneYee1966,Mendoza2010,Hauser2019}. The ELBM is advantageous to the FDTD method in its use of a single lattice, as opposed to an offset Yee-lattice \cite{KaneYee1966}. The Hauser \& Verhey (HV) ELBM has the additional capability of representing multi-dimensional propagation of electromagnetic waves which are free of numerical dispersion in free space \cite{Hauser2019}. Previously, both Chen \emph{et al.} and Hauser \& Verhey have used piece-wise linear recursive convolution (PLRC) methods to model the frequency dispersion of electromagnetic waves \cite{Chen2013,Hauser2021,hauser2022simulation}. The HV ELBM is second-order accurate, demonstrating an improvement over Chen \emph{et al.}'s ELBM which is first-order accurate \cite{Chen2013}, while remaining amenable to an ADE approach \cite{Warner2021}. Compared with the PLRC approach, which is typically used in the ELBM, an auxiliary differential equation (ADE) approach has demonstrated greater stability and efficiency for evaluating second-order accurate frequency dispersion of electromagnetic waves in the FDTD method \cite{Okoniewski1997,Chun2013}. The ADE with multiple complex-conjugate pole-residue pairs (CCPRPs)\cite{MinghuiHan2006} may also describe Debye and Lorentz media in a unified manner, reducing its implementation complexity from the PLRC method which treats these two types of media in an independent manner\cite{Chun2013,Chen2013}. In a previous conference paper \cite{Warner2021}, we integrated a second-order central difference of the ADE using multiple CCPRPs\cite{Okoniewski1997,MinghuiHan2006} within a one-dimensional (1D) HV ELBM\cite{Hauser2019} to model the broadband frequency-dispersion of arbitrary electromagnetic waves transmitted through dispersive silver \cite{MinghuiHan2006}. This study elaborates on our conference paper by describing how a central difference CCPRP ADE dispersive model designed for the FDTD method \cite{Okoniewski1997,MinghuiHan2006} can be incorporated into the HV ELBM \cite{Hauser2019} without modifying the ADE model or the HV ELBM. It further compares the 1D HV ELBM and the 1D FDTD method's relative errors, order-of-accuracy, computational time, and computational resources for evaluating the analytical transmittance of evaporated silver from Palik's measurements \cite{Lynch1985,MinghuiHan2006,Chen2013}. The relative error and computational time comparisons are drawn for the simplest case of steady-state continuous wave transmission through an evaporated silver slab of 100 \si{nm} thickness \cite{Lynch1985,MinghuiHan2006,Warner2021}. Finally, a Dirac-delta wave-function is used to demonstrate the stability of the 1D HV ELBM model, while recovering a broadband transmittance spectrum for the same evaporated silver slab of 100 \si{nm} thickness within a single simulation, which reduces the computational time by 2-3 orders of magnitude compared with the continuous-wave simulation.  Since electric and magnetic fields are synchronous in the 1D HV ELBM, we find it requires 40\% less computational time and only 10\% more memory for obtaining similar numerical solutions as the 1D FDTD method \cite{KaneYee1966,Hauser2019}. Therefore, including an ADE in the ELBM is a stable and accurate alternative for modeling broadband frequency dispersion of electromagnetic waves at optical frequencies.

\section{Theory}
This section describes the auxiliary differential equation (ADE) with added complex-conjugate pole-residue pairs (CCPRPs) for modeling dispersion of unified Debye and Lorentz media.

\subsection{Auxiliary Differential Equation}
The ADE\cite{Okoniewski1997} can be implemented using CCPRPs to model dispersion of unified Debye and Lorentz media \cite{MinghuiHan2006}.  The ADE with CCPRPs uses a dispersive permittivity, 
\begin{equation}\tag{1}\label{eq:1}
    \epsilon(\omega)=\epsilon_0\epsilon_\infty + \epsilon_0\sum_{p=1}^{N_p} \left(\frac{c_p}{j\omega-a_p}+\frac{c_p^*}{j\omega-a_p^*}\right),
\end{equation}
where $\omega$ is the angular frequency, $a_p$ and $c_p$ are the complex poles and residues, while $c_p^*$ and  $a_p^*$ are their conjugates \cite{MinghuiHan2006}. The dispersive permittivity $\epsilon(\omega\rightarrow\infty)=\epsilon_0\epsilon_\infty$, where $\epsilon_\infty$ is the relative permittivity of the medium as $\omega\rightarrow\infty$. The corresponding representations of $c_p$ and $a_p$ in terms of the pole relaxation time, $\tau_p$, plasmon frequency, $\omega_p$, and damping constant, $\delta_p$, are listed in Tab. \ref{tab:1}. 
\begin{table}[!t]
    \centering
   \caption{Representation of pole-pairs for Debye and Lorentz media according to Eq. (\ref{eq:2})}
    \begin{tabularx}{\textwidth}{| X | X |}
    \hline
    Debye medium parameters & Lorentz medium parameters\\
    \hline
       $c_p=\Delta \epsilon_p/(2\tau_p)$  &  $c_p=j\Delta\epsilon_p\omega_p^2/(2\sqrt{\omega_p^2-\delta_p^2})$ \\
        $a_p=-1/\tau_p$ &  $a_p=-\delta_p - j\sqrt{\omega_p^2-\delta_p^2}$ \\
        \hline
    \end{tabularx}
    \label{tab:1}
\end{table}
Treating Debye and Lorentz media in a unified manner generalizes the implementation of dispersion \cite{MinghuiHan2006}. In the frequency domain, the corresponding polarization current density for each CCPRP is defined as
\begin{equation}\tag{2}\label{eq:2}
    \begin{split}
    \mathbf{J}_p(\omega) = \epsilon_0\frac{c_p}{j\omega-a_p}j\omega \mathbf{E}(\omega), \\
        \mathbf{J}_p^*(\omega) = \epsilon_0\frac{c_p^*}{j\omega-a_p^*}j\omega \mathbf{E}(\omega),
    \end{split}
\end{equation}
where $\mathbf{E}(\omega)$ is the electric field intensity. In the time-domain, only real polarization current density is required, so the conjugate terms are redundant \cite{MinghuiHan2006}.
Therefore, the auxiliary differential equation (ADE) for each CCPRP's current density in the time-domain becomes
\begin{equation}\tag{3}\label{eq:3}
    \frac{d }{dt} \mathbf{J}_p(\mathbf{x},t) - a_p\mathbf{J}_p(\mathbf{x},t) - \epsilon_0c_p\frac{d}{dt}\mathbf{E}(\mathbf{x},t)=0.
\end{equation}
This ADE can be implemented for multiple CCPRPs in both the finite-difference time-domain (FDTD) method and the Hauser \& Verhey electrodynamic lattice-Boltzmann method (HV ELBM).
\section{Methodology}
We compare the HV ELBM to the FDTD method for modeling one-dimensional frequency dispersion of electromagnetic waves transmitted through a 100 \si{nm} silver (Ag) slab \cite{Hauser2019,MinghuiHan2006}. This section first discusses the implementation of a central-difference ADE to describe current density in the HV ELBM \cite{MinghuiHan2006,Okoniewski1997,Hauser2019}. Then this section describes the CCPRP model used to represent silver \cite{MinghuiHan2006}, and concludes with a description of the methodology used to compare the HV ELBM with the same CCPR model used in the FDTD method \cite{MinghuiHan2006,Okoniewski1997}.
\subsection{Electrodynamic lattice-Boltzmann method}
The HV ELBM is implemented on a one-dimensional three-velocity (D1Q3) lattice which can incorporate $N=4$ independent non-zero left-handed and right-handed orthogonal bases with velocity, $\mathbf{c}_n$, electric field, $\mathbf{e}_n$, and magnetic field, $\mathbf{h}_n$, lattice vectors that satisfy a Chapman-Enskog analysis \cite{Hauser2019}.  In one dimension there is no orthogonal propagation and the sets of lattice vectors reduce to scalar sequences constrained in their respective dimensions. The sequences for the D1Q3 lattice used in this work are
\begin{align}
     &  \mathbf{c}_{n} = c_{n}\hat{\mathbf{x}} = \{1 \quad \mbox{-}1 \quad \mbox{-}1 \quad 1\}\hat{\mathbf{x}}, \tag{4a}\label{eq:4a} \\
     & \mathbf{e}_n =  e_{n}\hat{\mathbf{y}} = \{1 \quad 1 \quad \mbox{-}1 \quad \mbox{-}1\}\hat{\mathbf{y}}, \:\:\textrm{and} \tag{4b}\label{eq:4b} \\
     &  \mathbf{h}_n = h_{n}\hat{\mathbf{z}} = \{1 \quad \mbox{-}1 \quad 1 \quad \mbox{-}1\}\hat{\mathbf{z}}. \tag{4c}\label{eq:4c}
\end{align}
Given that there is no orthogonal propagation in one dimension, we find that the electric and magnetic fields propagate one lattice unit cell per each iteration using the sequences (\ref{eq:4a}-\ref{eq:4c}), such that the lattice unit is $\delta x=1$ and the time unit is $\delta t=1$, to represent the lattice velocity $c_0=1$. With a greater number of dimensions, orthogonal outcomes are required; therefore $\delta t=1/2$ for $\delta x=1$ and $c_0=1$ \cite{Hauser2019}. 
 In one-dimension, we find that the 1/2 scaling factor for Hauser \& Verhey's Chapman-Enskog analysis for Maxwell's equations manifests exclusively within the initial condition while expected wave behaviour occurs at each consecutive time-step. The corresponding equilibrium pseudo-particle population is
\begin{equation} \tag{5}\label{eq:5}
    f_n^{eq}(\mathbf{x},t) = \frac{1}{4}(\mathbf{E}(\mathbf{x},t)\cdot\mathbf{e}_n+\mathbf{H}(\mathbf{x},t)\cdot\mathbf{h}_n),
\end{equation}
while the relaxation time is $\tau=1/2$. Therefore, the collision operation is
\begin{equation} \tag{6}\label{eq:6}
   f_n(\mathbf{x}+\mathbf{c}_n\delta t,t+\delta t) = 2f_n^{eq}(\mathbf{x},t)-f_n(\mathbf{x},t),
\end{equation}
where $f_n(\mathbf{x},t)$ is outcome $n$ of the pseudo-particle population. 
\\
\subsubsection{Non-dispersive dielectric media}
The equilibrium polarization density, $\mathbf{P}^{eq}(\mathbf{x},t)$, is the outcome of the electric field remaining at its present location. This outcome is not included within $f^{eq}(\mathbf{x},t)$, but is crucial for electric and magnetic field interactions with dielectric media \cite{Hauser2017}. In a uniform non-dispersive medium, $\mathbf{P}^{eq}(\mathbf{x},t)$ can be defined as \cite{Hauser2017,Hauser2019}
\begin{equation}\tag{7}\label{eq:7}
    \mathbf{P}^{eq}(\mathbf{x},t)=(\epsilon_r(\mathbf{x})-1)\mathbf{E}(\mathbf{x},t),
\end{equation}
while the dynamic polarization density, $\mathbf{P}(\mathbf{x},t)$, is calculated with a collision operation,
\begin{equation}\tag{8}\label{eq:8}
    \mathbf{P}(\mathbf{x},t+\delta t)=2\mathbf{P}^{eq}(\mathbf{x},t)-\mathbf{P}(\mathbf{x},t).
\end{equation}
The polarization density is included within a re-normalized electric field, at the same location and time instant as the magnetic field. In this way, the electric and magnetic fields are calculated as synchronous orthogonal moments of $f(\mathbf{x},t)$,
\begin{equation}\tag{9}\label{eq:9}
\begin{split}
    & \mathbf{E}(\mathbf{x},t) = \frac{1}{\epsilon_r(\mathbf{x})}\left(\sum_{n=1}^Nf_n(\mathbf{x},t)\mathbf{e}_n+\mathbf{P}(\mathbf{x},t)\right) \\
   & \mathbf{H}(\mathbf{x},t) = \sum_{n=1}^Nf_n(\mathbf{x},t)\mathbf{h}_n.
    \end{split}
\end{equation}
\subsubsection{Dispersive dielectric media}
To model dispersive media using the HV ELBM, we include loss in the equilibrium condition for the non-dispersive polarization density, $\mathbf{P}^{eq}(\mathbf{x},t)$, in (\ref{eq:7}) via the Bhatnaghar-Gross-Krook (BGK) collision operator \cite{Bhatnagar1954}. This approach is analogous to the temporal implicit central difference scheme of the ADE used in the FDTD method \cite{Okoniewski1997,MinghuiHan2006}, and the Shan-Chen forcing scheme used in the kinetic lattice-Boltzmann method \cite{Shan1993}. It is also similar to the finite-element scheme used by Hauser \& Verhey for a uniform current density \cite{Hauser2019}. First the CCPRP parameters are converted to dimensionless forms as $\alpha_p=a_p(\Delta t e/\hbar)$, and $\chi_p = c_p(\Delta t e/\hbar)$, using  $\Delta t$ as the time-step, $e$ as the elementary unit of charge, and $\hbar$ as the reduced Planck constant for conversion from angular- to time-frequency \cite{MinghuiHan2006}. We interpret each CCPRP polarization current density in the ADE of  (\ref{eq:3}) as a population of the collision operator for the polarization density \cite{Hauser2021,hauser2022simulation}. Written in dimensionless form,
\begin{equation}\tag{10}\label{eq:10}
\begin{split}
    \mathbf{\Omega}_p(\mathbf{P}_p) & = -\mathbf{j}_p(\mathbf{x},t)=\frac{\chi_p}{\alpha_p}\frac{d}{dt} \mathbf{E}(\mathbf{x},t)-\frac{1}{\alpha_p}\frac{d}{dt} \mathbf{j}_p(\mathbf{x},t), \\
    & = -\frac{d}{dt}\mathbf{P}_p(\mathbf{x},t)=-\frac{\delta t}{\tau}(\mathbf{P}_p(\mathbf{x},t)-\mathbf{P}_p^{eq}(\mathbf{x},t)),
     \end{split}
\end{equation}
where $\mathbf{\Omega}_p(\mathbf{P}_p)$ is a population of the collision operator, which we propose takes a BGK form consistent with non-dispersive media \cite{Hanasoge2011,Hauser2017,Hauser2019}, as a function of the dimensionless polarization density for each CCPRP, $\mathbf{P}_p(\mathbf{x},t)$. We use a conventional backward difference in time to estimate each dimensionless CCPRP polarization current density in the present iteration
\begin{equation}\tag{11}\label{eq:11}
    \mathbf{j}_p( \mathbf{x},t)= \kappa_p  \mathbf{j}_p(\mathbf{x},t-\delta t)+\frac{\beta_p}{\delta t} (\mathbf{E}(\mathbf{x},t)-\mathbf{E}(\mathbf{x},t-\delta t)),
\end{equation}
where $k_p$ and $\beta_p$ are defined as the dimensionless parameters 
\begin{equation}\tag{12}\label{eq:12}
    \kappa_p = \frac{1+\alpha_p\tau}{1-\alpha_p\tau}\qquad \beta_p = \frac{\chi_p}{1-\alpha_p\tau},
\end{equation}
which are identical to their formulation in the FDTD method \cite{MinghuiHan2006}. In higher dimensions, additional considerations must be made to preserve Gauss's law \cite{Hauser2019}, and more pole-pairs to preserve the model's stability \cite{Hauser2021,hauser2022simulation}.  Sixteen-pole Debye and Lorentz models for dispersive media have been described elsewhere \cite{Hauser2021,hauser2022simulation}.  
Now a dispersive equilibrium polarization density, $\mathbf{j}^{eq}(\mathbf{x},t)$, is constructed as the sum of $p=\{1,...,N_p\}$ finite-difference approximations for each $\mathbf{j}_p(\mathbf{x},t)$ in a similar manner to its evaluation in the FDTD method \cite{MinghuiHan2006}. However, we re-interpret their sum as a moment of the BGK CCPRP collision operator, $\mathbf{\Omega}_p(\mathbf{P}_p)$, from (\ref{eq:10})
 \begin{equation}\tag{13}\label{eq:13}
     \begin{split}
\mathbf{j}^{eq}(\mathbf{x},t) & =\textrm{Re}\left\{\sum_{p=1}^{N_p}\mathbf{j}_p(\mathbf{x},t)(1+\kappa_p)\right\}, \\
& = \textrm{Re}\left\{\sum_{p=1}^{N_p}\mathbf{\Omega}_p(\mathbf{P}_p)(1+\kappa_p)\right\}, \\
& = \frac{\delta t}{\tau}(\mathbf{P}(\mathbf{x},t)-\mathbf{P}^{eq}(\mathbf{x},t)).
 \end{split}
 \end{equation}
 In an earlier work by Hauser \& Verhey \cite{Hauser2017}, the update equation for the polarization density was described using the BGK collision operator
 \begin{equation}\tag{14}\label{eq:14}
    \begin{split}
     \mathbf{P}(\mathbf{x},t+\delta t) & =\mathbf{P}(\mathbf{x},t)-\frac{\delta t}{\tau}(\mathbf{P}(\mathbf{x},t)-\mathbf{P}^{eq}(\mathbf{x},t)), \\
     & = \mathbf{P}(\mathbf{x},t)-\mathbf{j}^{eq}(\mathbf{x},t).
     \end{split}
 \end{equation}
 Therefore, it must also be true that
 \begin{equation}\tag{15}\label{eq:15}
    \mathbf{j}^{eq}(\mathbf{x},t)= -\left(\mathbf{P}(\mathbf{x},t+\delta t) -\mathbf{P}(\mathbf{x},t)\right).
 \end{equation}
 We can rearrange the collision operator from (\ref{eq:14}) to reveal the second-order equilibrium polarization density in terms of the equilibrium current density 
 \begin{equation}\tag{16}\label{eq:16}
 \begin{split}
     \mathbf{P}^{eq}(\mathbf{x},t) & =\mathbf{P}(\mathbf{x},t)+\frac{\tau}{\delta t}(\mathbf{P}(\mathbf{x},t+\delta t)-\mathbf{P}(\mathbf{x},t)) \\
     & =\mathbf{P}(\mathbf{x},t)-\frac{\tau}{\delta t}\mathbf{j}^{eq}(\mathbf{x},t).
     \end{split}
 \end{equation}
This new equilibrium population for the polarization density, $\mathbf{P}^{eq}(\mathbf{x},t)$, can be adopted without modifying the original HV ELBM formulation \cite{Hauser2019}. However, to enforce its stability we further subject the dynamic polarization density, $\mathbf{P}(\mathbf{x},t)$, to the constraint that it must relax to a state of non-dispersive equilibrium from (\ref{eq:7}), i.e. $\mathbf{P}(\mathbf{x},t)\rightarrow\mathbf{P}^{eq}(\mathbf{x},t)=(\epsilon_r(\mathbf{x})-1)\mathbf{E}(\mathbf{x},t)$, within one time unit.
 Given that the relaxation time is constrained to $\tau=1/2$ for elastic scattering \cite{Hanasoge2011}, we simply subtract the effective equilibrium polarization current density from (\ref{eq:7}) to represent dispersion of the equilibrium polarization density,
\begin{equation}\tag{17}\label{eq:17}
    \mathbf{P}^{eq}(\mathbf{x},t)=(\epsilon_r(\mathbf{x})-1)\mathbf{E}(\mathbf{x},t)-\frac{\tau}{\delta t}\mathbf{j}^{eq}(\mathbf{x},t),
\end{equation}
where
\begin{equation}\tag{18}\label{eq:18}
 \epsilon_r = \epsilon_\infty + \sum_{p=1}^{N_p} \textrm{Re} \{ \beta_p \},
\end{equation}
represents the non-dispersive relative permittivity of the CCPRPs at equilibrium \cite{MinghuiHan2006}. Following this calculation of the equilibrium polarization density, $\mathbf{P}(\mathbf{x},t)$ can be evaluated according to (\ref{eq:8}) and $\mathbf{E}(\mathbf{x},t)$ and $\mathbf{H}(\mathbf{x},t)$ according to (\ref{eq:9}) without modifying the HV ELBM formulation \cite{Hauser2019}. Further work is required to show that this ADE approach agrees with analytical solutions in higher dimensions. This work employs a one dimensional analysis to demonstrate the model's unconditional stability and recovery of broadband transmittance, which is useful for quick frequency domain simulations.  

\subsection{The discrete Dirac-delta wave-function} 
The 1D HV ELBM is unconditionally stable, and may therefore represent the discrete Dirac-delta wave-function as its initial condition. The discrete Dirac-delta wave-function has a uniform frequency spectrum up to the spatiotemporal Nyquist limit in one dimension, making it the ideal source condition for modeling broadband electromagnetic frequency dispersion. Its initial condition is an electric and magnetic field represented at a single location in space, $\mathbf{x}_0$, for a single instant in time, $t_0$,
\begin{equation}\tag{19}\label{eq:19}
    \mathbf{E}(\mathbf{x}_0,t_0)=\frac{1}{2\delta x}\mathbf{e}_n, \qquad
    \mathbf{H}(\mathbf{x}_0,t_0) = \frac{1}{2\eta \delta x}\mathbf{h}_n.
\end{equation}
The $1/2$ factor manifests exclusively within the initial condition as a scaling factor of Hauser \& Verhey's Chapman-Enskog analysis to Maxwell's equations \cite{Hauser2019}. This is expected, given the duplicate elements in $\mathbf{e}_n$ and $\mathbf{h}_n$ that are required to describe all configurations of the electric field, the magnetic field and velocity on the lattice. Initializing one outcome with a half-amplitude is necessary to represent a unit amplitude, which is maintained after the first iteration in the 1D HV ELBM. Alternatively, the 1D FDTD method is conditionally stable up to the Courant-Friedrichs-Lewy (CFL) limit \cite{Courant1928,KaneYee1966,Sarris2011}, such that it is not possible for the 1D FDTD method to represent the discrete Dirac-delta wave-function and its frequency components at the Nyquist limit. The 1D FDTD method will instead become unstable and exhibit unbounded growth in numerical error.
\subsection{Power spectral density} 
The power spectral density is recovered from the Poynting vector of electromagnetic sine waves fully transmitted through the 100 \si{nm} silver slab at multiple frequencies. We believe this is the most intuitive way to evaluate and compare the relative error of the simulated dispersion in both the HV ELBM and the FDTD method. One advantage of the 1D HV ELBM over the 1D FDTD method is that the 1D HV ELBM can represent a Dirac-delta wave-function with a uniform power spectral density approaching the Nyquist limit. Given this input, we demonstrate the accurate recovery of the 100 \si{nm} silver slab's transmittance as the multiplication of fast-Fourier transforms (FFTs) of electric and magnetic fields fully transmitted through the slab. Their multiplication in the frequency domain is equivalent to their convolution in the time-domain,
\begin{equation}\tag{20}\label{eq:20}
    \mathbf{\hat{S}}(\nu)=\mathbf{\hat{E}}(\nu)\times\mathbf{\hat{H}}(\nu),
\end{equation}
where $\nu=\Delta t/T'$ represents the dimensionless time-frequency for a dimensional (physical) time unit $\Delta t$ and time period $T'$. The power spectral density of the Dirac-delta wave-function's fully transmitted electric and magnetic fields approximates the transmittance spectrum of the 100 \si{nm} silver slab as an impulse response function with an $l$\textsuperscript{th} order-of-accuracy, $\mathcal{O}(\Delta \nu^l)$,
\begin{equation}\tag{21}\label{eq:21}
    T_{slab}(\nu)=|\mathbf{\hat{S}}(\nu)|+\mathcal{O}(\Delta \nu^l).
\end{equation}
In this manner, one simulation can be used to retrieve a broadband transmittance spectrum representing the frequency-dispersion of electromagnetic waves through a dispersive medium, as opposed to the inherently band-limited spectra that represent all continuous wave-functions. The error order-of-accuracy is quantified using the first-order, $L_1$, and second-order, $L_2$, relative error norms as a function of the number of photon energy bins, $N_\nu$, across the spectrum. The first-order and second-order norms are defined as
\begin{equation}\label{eq:22}\tag{22}
\begin{aligned}
   & L_1=\frac{1}{N_\nu}\sum_{\nu}^{N_\nu}\left|\frac{T_{slab}(\nu)-|\hat{\mathbf{S}}(\nu)|}{T_{slab}(\nu)}\right|, \\
   & L_2=\left(\frac{\sum_{\nu}^{N_\nu}|T_{slab}(\nu)-|\hat{\mathbf{S}}(\nu)||^2}{\sum_{\nu}^{N_\nu}|T_{slab}(\nu)|^2}\right)^{1/2}.
\end{aligned}
\end{equation}
\subsection{Material Model}
\begin{figure}[!t]
    \centering
    \includegraphics[width=\textwidth]{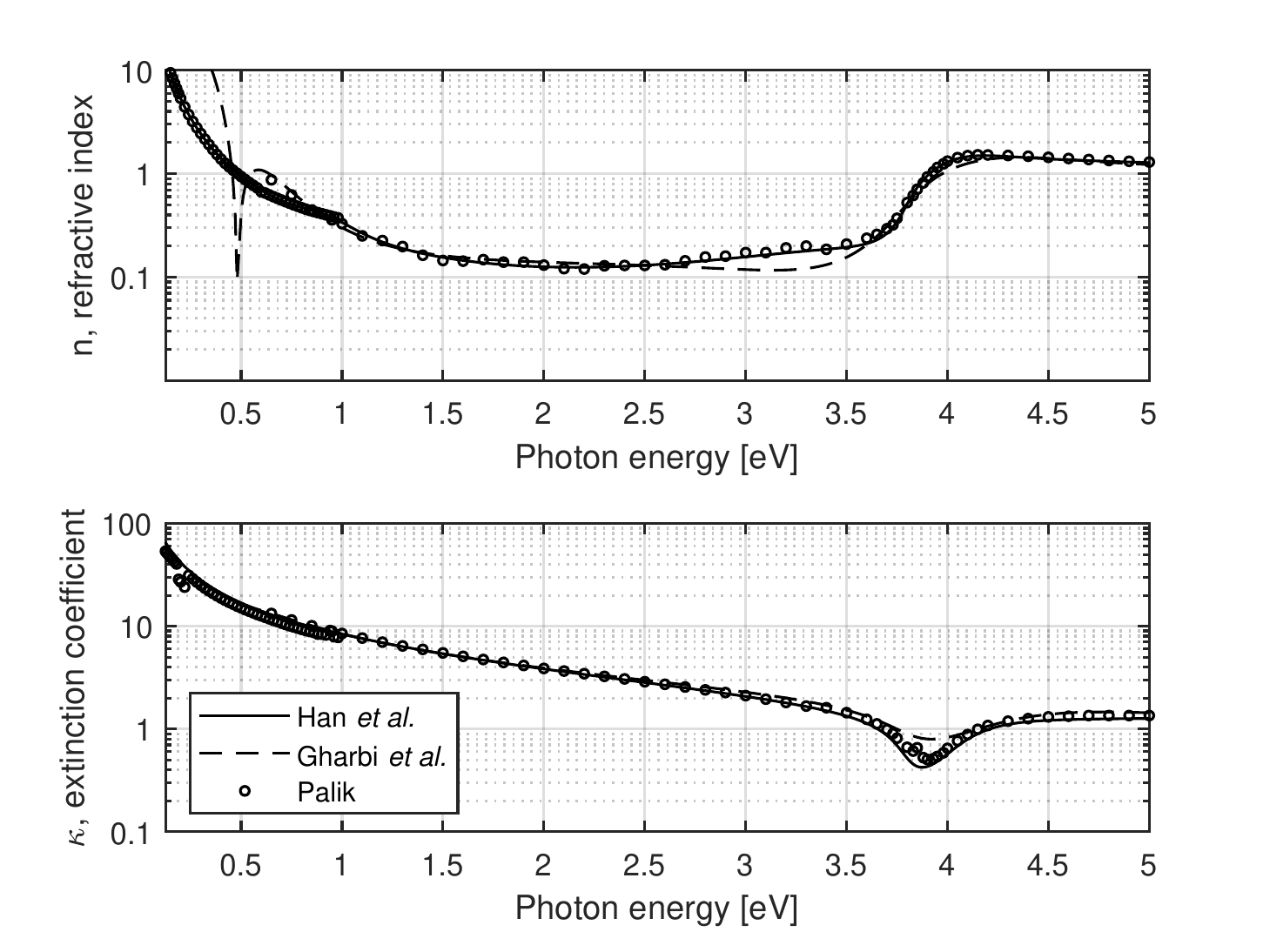}
    \caption{Comparison of Han \emph{et al.} \cite{MinghuiHan2006}, and Gharbi \emph{et al.} \cite{Gharbi2020} fitting to measured index of refraction, $n$, and extinction coefficient, $k$, for evaporated silver (Ag) \cite{Lynch1985}. Han \emph{et al.}'s model provides a better fit.}
    \label{fig:1}
\end{figure}

We choose a CCPRP model which represents dispersive silver (Ag) using six CCPRPs \cite{MinghuiHan2006}. We compare this model to measurements of the refractive index, $n$, and extinction coefficient, $k$, for evaporated silver \cite{Lynch1985}, and a more recent fit obtained using four CCPRPs \cite{Gharbi2020}. We find that six CCPRPs, including $a_p$ and $b_p$, can more accurately represent the measurement data for evaporated silver \cite{MinghuiHan2006,Lynch1985}. Tab. \ref{tab:2} lists these parameters, and Fig. \ref{fig:1} demonstrates their fit to empirical data. Nonetheless, we find that this CCPRP model has insufficient pole-pairs to accurately represent the available bandwidth of the 1D HV ELBM, so we only compare its error with the FDTD method in the photon energy range of interest, 1-5 \si{eV}.
\begin{table}[!t]
    \centering
        \caption{Values of the Complex-Conjugate Pole-Residue Pairs for the permittivity of Ag\cite{MinghuiHan2006}}
    \begin{tabularx}{\textwidth}{|c X|c X|}
    \hline
    $c_p$, & \si{eV} & $a_p$, & \si{eV} \\
    \hline
       \scriptsize $c_1=$ & \scriptsize $5.987\cdot10^{\mbox{-}1}+\textrm{j}4.195\cdot 10^3$ & \scriptsize $a_1=$ & \scriptsize $\mbox{-}2.502\cdot10^{\mbox{-}2} - \textrm{j}8.626\cdot10^{\mbox{-}3}$   \\
        \scriptsize$c_2=$ & \scriptsize$\mbox{-}2.211\cdot10^{\mbox{-}1}+\textrm{j}2.680\cdot10^{\mbox{-}1}$ &\scriptsize $a_2=$ &\scriptsize $\mbox{-}2.021\cdot10^{\mbox{-}1} - \textrm{j}9.407\cdot10^{\mbox{-}1}$  \\
        \scriptsize$c_3=$ &\scriptsize  $\mbox{-}4.240+\textrm{j}7.324\cdot10^{\mbox{-}2}$ &\scriptsize $a_3=$ &\scriptsize $\mbox{-}1.467\cdot10^1 - \textrm{j}1.338$ \\
        \scriptsize$c_4=$ &\scriptsize $6.391\cdot10^{\mbox{-}1}-\textrm{j}7.186\cdot10^{\mbox{-}2}$ &\scriptsize $a_4=$ &\scriptsize $\mbox{-}2.997\cdot10^{\mbox{-}1}-\textrm{j}4.034$ \\
       \scriptsize $c_5=$ &\scriptsize $1.806 + \textrm{j}4.563$ &\scriptsize $a_5=$ &\scriptsize $\mbox{-}1.896-\textrm{j}4.808$ \\
       \scriptsize $c_6=$ &\scriptsize $1.443-\textrm{j}8.129\cdot10^1$ &\scriptsize $a_6=$ &\scriptsize $\mbox{-}9.396-\textrm{j}6.477$ \\
        \hline
    \end{tabularx}
    \label{tab:2}
\end{table}
\subsection{Simulation}\label{sec:sim}
 Three different source conditions are used to analyze the accuracy of electromagnetic wave frequency-dispersion from a 100 \si{nm} silver slab in the photon energy range of interest (1-5 \si{eV}): (1) a Ricker wavelet with a peak photon energy of $\hbar\omega_p/e=3.8735$ \si{eV} and a half-breadth of $\tau_b=145.32$ \si{as} (7.02 time units); (2) a sequence of time-harmonic sine-waves of variable period; and (3) a Dirac-delta wave-function.
 Source condition (1) is used to compare the numerically solved electric and magnetic fields in the 1D FDTD method and the 1D HV ELBM. Source condition (2) is used to compare their numerical accuracy and computational time. Source condition (3) is used to demonstrate the versatility of the 1D HV ELBM for modeling single-simulation broadband frequency dispersion of electromagnetic waves. \\ \\  
 Source condition (1) is situated at node $x_0=125$ on a lattice consisting of $N_x=500$ spatial unit cells. The silver slab is situated at the center of the lattice, and its thickness is 99.2 \si{nm}, to represent an integer number of dimensional spatial units each of 1/40th the smallest wavelength of interest ($\Delta x=6.2$ \si{nm}). This 3100 \si{nm} spatial domain is identical to that of a former FDTD simulation \cite{MinghuiHan2006}, to ensure a faithful reproduction of the original numerical solution. \\ \\ 
 Source condition (2) is situated at position $x_0=N_{x_k}/4$ within a 620 \si{nm}  spatial domain. The spatial domain's dimensional length is kept constant, but the number of its lattice units is linearly increased for each consecutive simulation indexed by $k$. The $k^{\textrm{th}}$ simulation has $N_{x_k}=100k$ spatial lattice units and represents $N_{\nu_k}=16k$ frequencies and $16k$ points of the transmittance spectrum. Each simulation achieves a steady state and the time-averaged Poynting vector is measured to estimate the numerical transmittance. As the number of spatial units increases, the silver slab thickness approaches 100 \si{nm}. The silver slab's exact thickness (from the number of spatial lattice units) is used in the analytical solution to assess the relative error norms. The 1D HV ELBM and 1D FDTD method relative error norms are plotted with respect to $N_\nu$, and given a similar accuracy, their computational time and memory are compared via linear regression.\\ \\
Source condition (3) is first initialized at dimensionless time $t_0=0$ and position $x_0=N_{x}/4$ in a 3100 \si{nm} spatial domain to illustrate the stable transmission of electromagnetic waves through the 100 \si{nm} slab in the 1D HV ELBM with the added ADE. Source condition (3) is then initialized at dimensionless time $t_0=0$ and position $x_0=N_{x_k}/4$ in a 620 \si{nm} spatial domain used for a sequence of $k$ simulations with $N_{x_k}=100k$ lattice units and $N_{t_k}=4000k$ time units. The transmitted power spectral density from source condition (3) is compared with the analytical transmittance spectrum of the 100 \si{nm} silver slab. The power spectral density's relative error (relative difference from the analytical transmittance spectrum), for source (3) is plotted with respect to the number of unit cells to determine the order-of-accuracy of the 1D HV ELBM solutions from the slope of a proportional error norm.

\subsection{Boundary conditions}
Absorbing boundary conditions are used on each end of the one dimensional spatial domain described in subsection \ref{sec:sim}, but with a distinct implementation in the FDTD and the HV ELBM. We choose the Mur boundary condition for the FDTD simulation, since it perfectly absorbs electric and magnetic fields in one dimension \cite{mur1981absorbing}. In the HV ELBM simulation we implement similar absorbing boundary conditions, called ``free'' boundaries, which assign the  boundary particle populations to their nearest adjacent nodes inside the domain,
\begin{equation}\tag{23}\label{eq:23}
    f_n(\mathbf{x}_b-\delta x\hat{\mathbf{n}}_b,t)=f_n(\mathbf{x}_b,t),
\end{equation}
where $\hat{\mathbf{n}}_b$ is the unit vector normal to boundary $b$ at position $\mathbf{x}_b$.
\newpage

\section{Results and Discussion}
\begin{figure} [H]
    \centering
    \includegraphics[width=\textwidth]{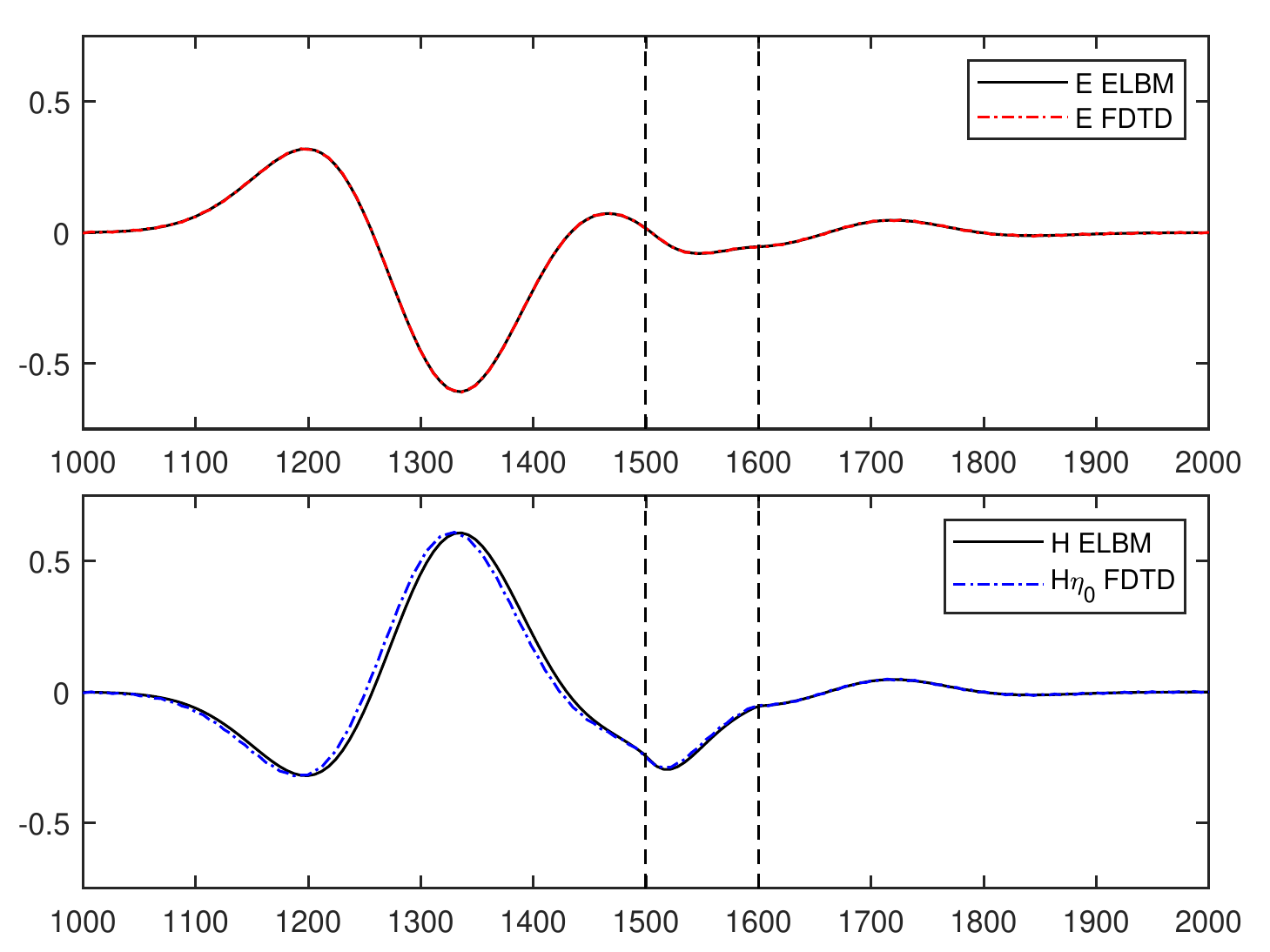}
    \caption{[Color online]: The electric field (top) and magnetic field (bottom) for the transmission of a Ricker wavelet with a peak photon energy of $\hbar\omega_p/e=3.8735$ \si{eV} and a half-breadth of $\tau_b=145.32$ \si{as} (7.02 time units) through a silver slab of 100 \si{nm} thickness (boundaries represented by black-dashed lines). Both the 1D FDTD method and the 1D HV ELBM recover similar electric and magnetic fields, apart from a phase delay in the magnetic field as a consequence of the Yee-lattice. Axis units are in nanometers, \si{nm}.}
    \label{fig:2}
\end{figure}

Fig. \ref{fig:2} illustrates a Ricker wavelet with a peak photon energy of $\hbar\omega_p/e=3.8735$ \si{eV} and a half-breadth of $\tau_b=145.32$ \si{as} (7.02 time units) as it transmits through a 100 \si{nm} thick silver slab in both the 1D FDTD method and the 1D HV ELBM. The 1D HV ELBM and 1D FDTD method predict similar incident, reflected, and transmitted electric and magnetic fields. However, there is a slight difference in phase between the magnetic field evaluated in the 1D HV ELBM and 1D FDTD method. This slight difference in phase may be attributed to the staggered Yee-lattice in the FDTD method, which will evaluate the magnetic field at a different time and location in space than the fully synchronous 1D HV ELBM.

\begin{figure}[!t]
    \centering
    \subfloat[]{\includegraphics[width=0.49\textwidth]{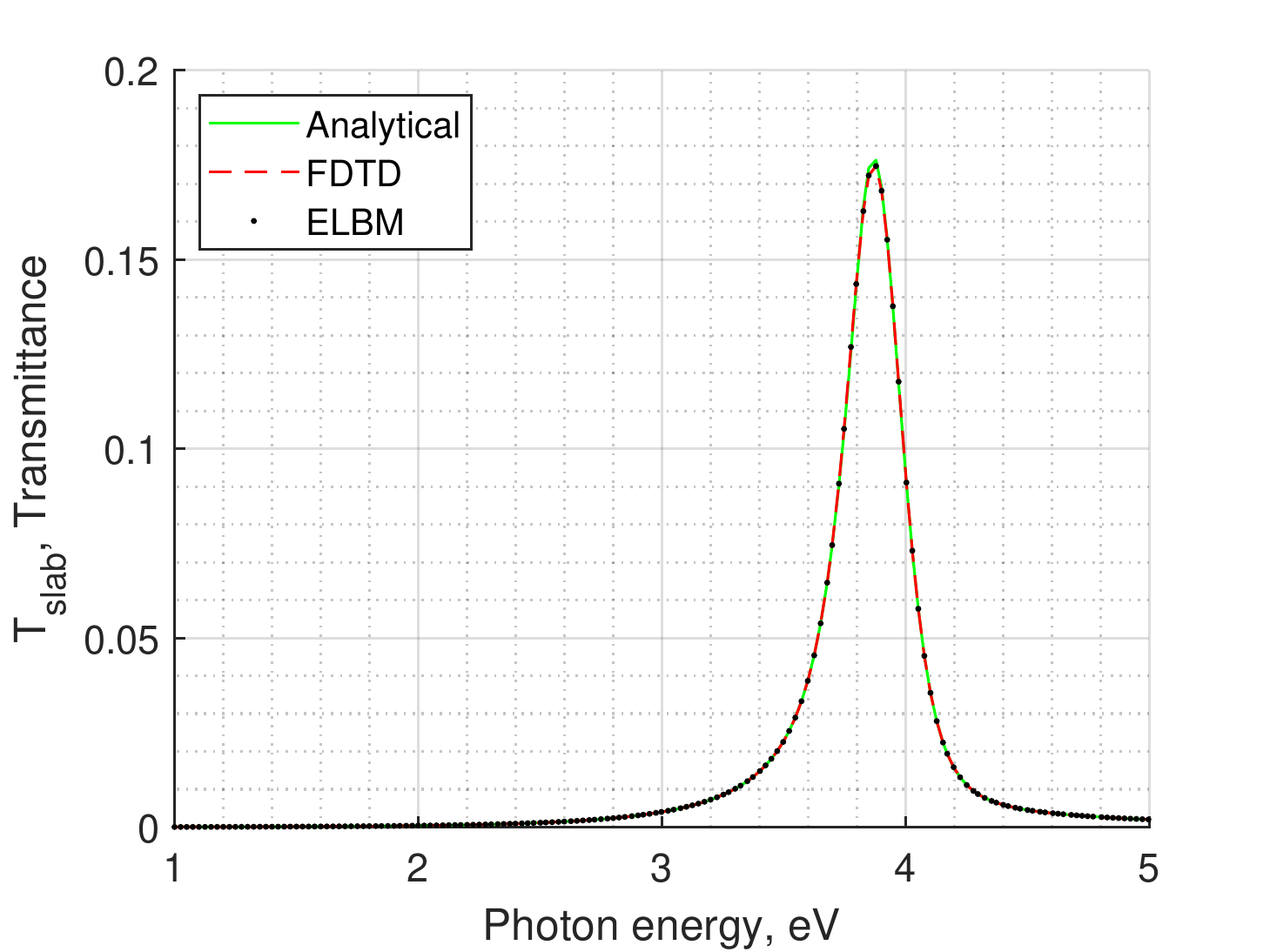}\label{fig:3a}}
     \subfloat[]{\includegraphics[width=0.49\textwidth]{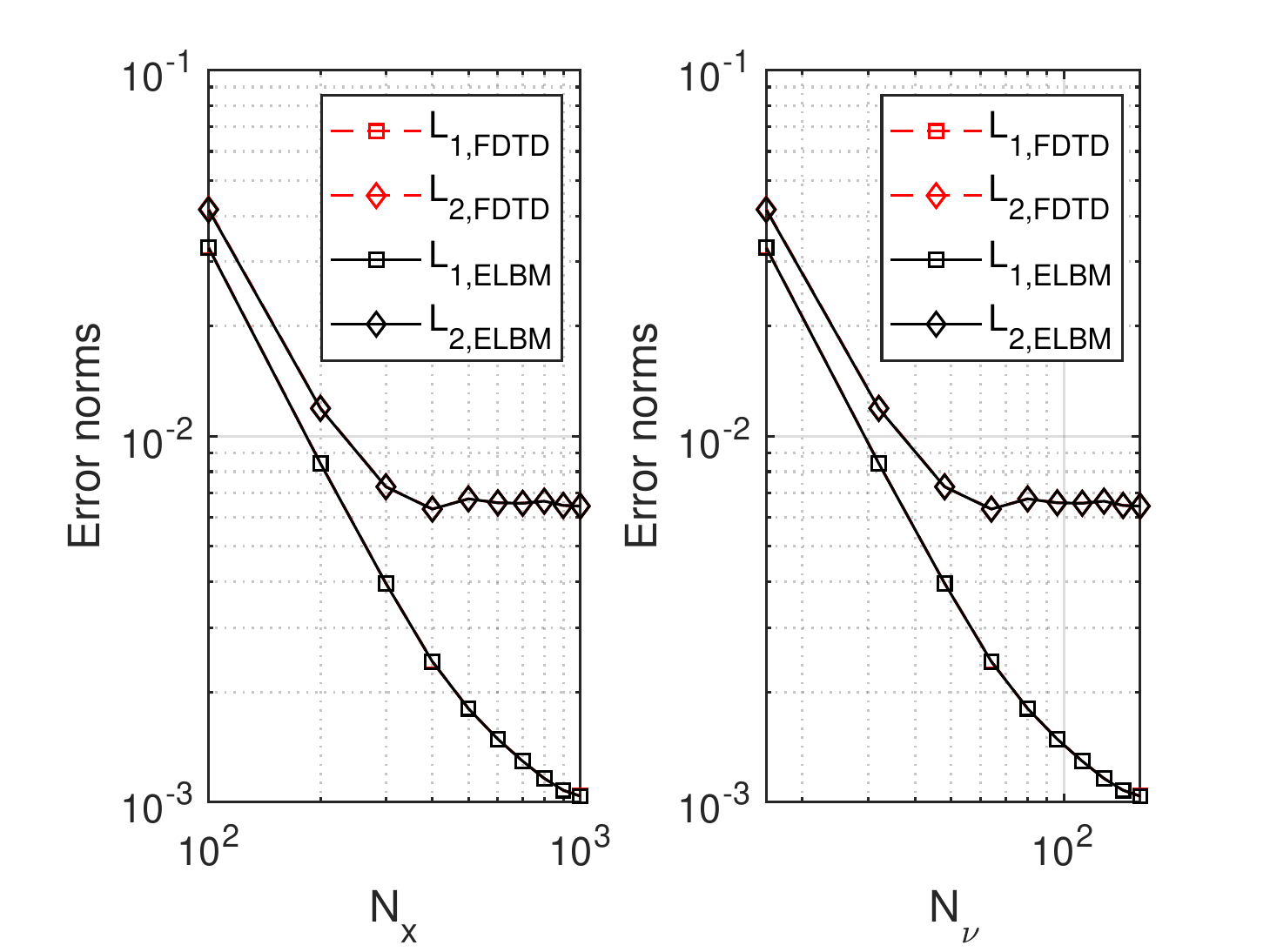}\label{fig:3b}}
    \caption{[Color online]: (a) The transmittance spectrum from 160 continuous waves [source condition (2)] in the photon energy range 1-5 \si{eV} transmitted through a 100 \si{nm} silver slab. (b) The error norms ($L_1$) and ($L_2$) from the transmittance spectrum measured for an increasing number of spatial lattice units ($N_x$) and frequencies ($N_{\nu}$) between 1-5 \si{eV}. Red symbols are beneath the black symbols, both solutions have the same numerical error.}
    \label{fig:3}
\end{figure}

\begin{figure}[!t]
    \centering
    \subfloat[]{\includegraphics[width=0.49\textwidth]{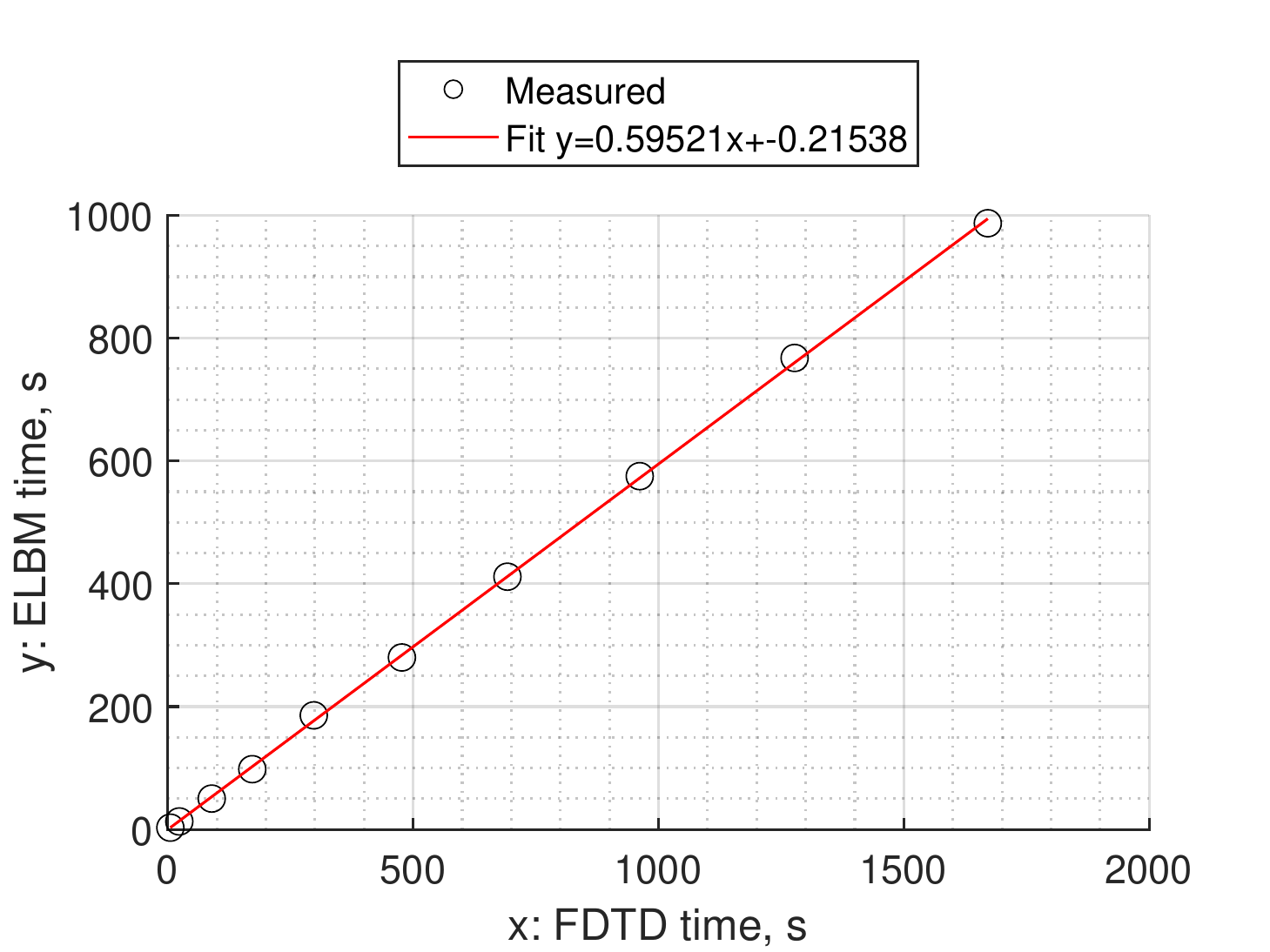}\label{fig:4a}}
     \subfloat[]{\includegraphics[width=0.49\textwidth]{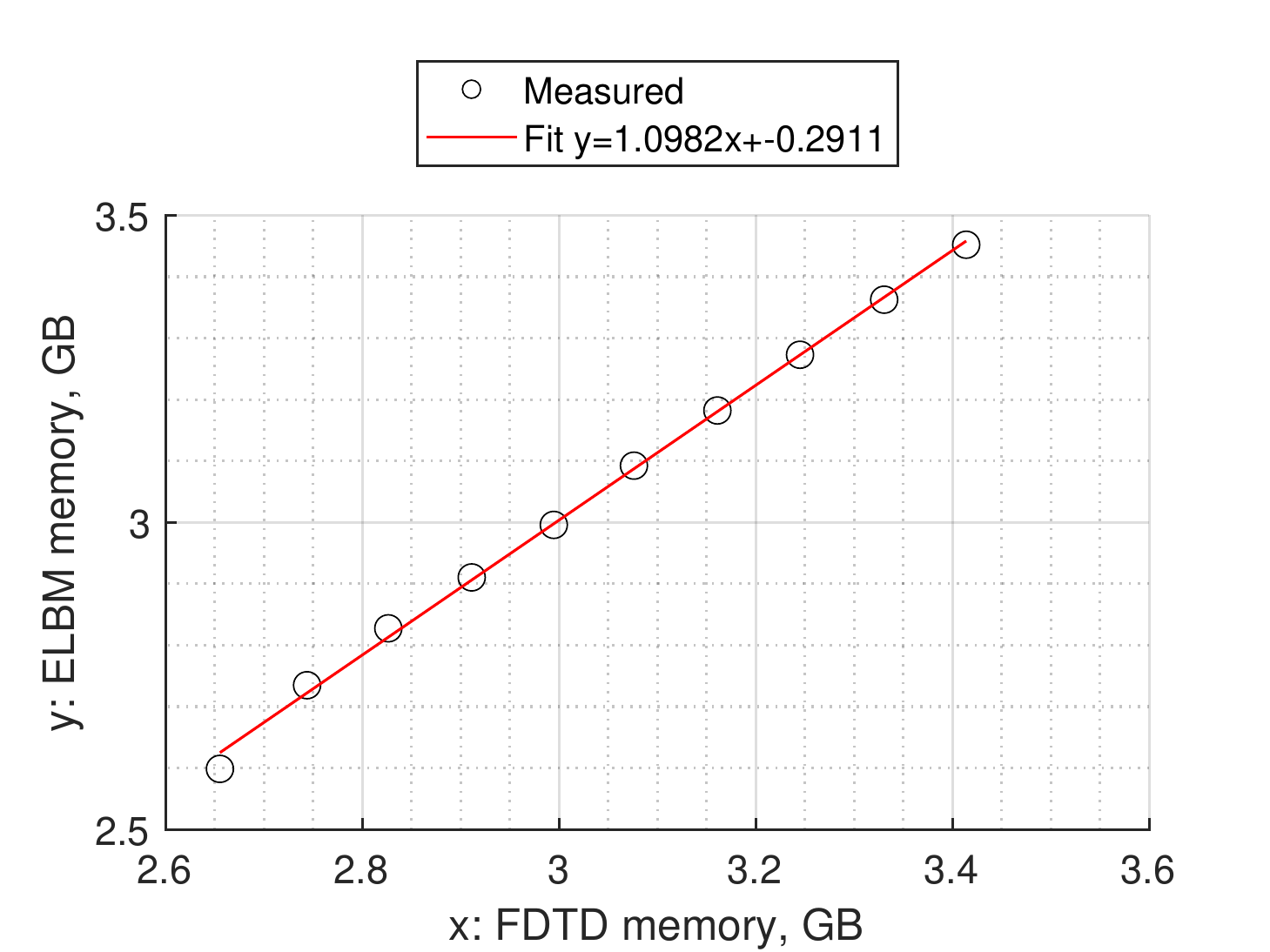}\label{fig:4b}}
    \caption{[Color online]: (a) The computational time and (b) memory for the simulation of multiple continuous waves [source condition (2)]. See \ed{Supplementary material} for MATLAB scripts. The 1D HV ELBM requires 60\% of the computation time, but 10\% greater memory to achieve the same numerical solutions from 1-5 \si{eV}. Simulations were performed on an 11\textsuperscript{th} Gen Intel\textsuperscript{\textregistered} Core\textsuperscript{TM} i7-11700K @ 3.60 GHz processor with 32 GB of RAM and 8 cores. }
    \label{fig:4}
\end{figure}

\begin{figure} [t!]
    \centering
    \includegraphics[width=\textwidth]{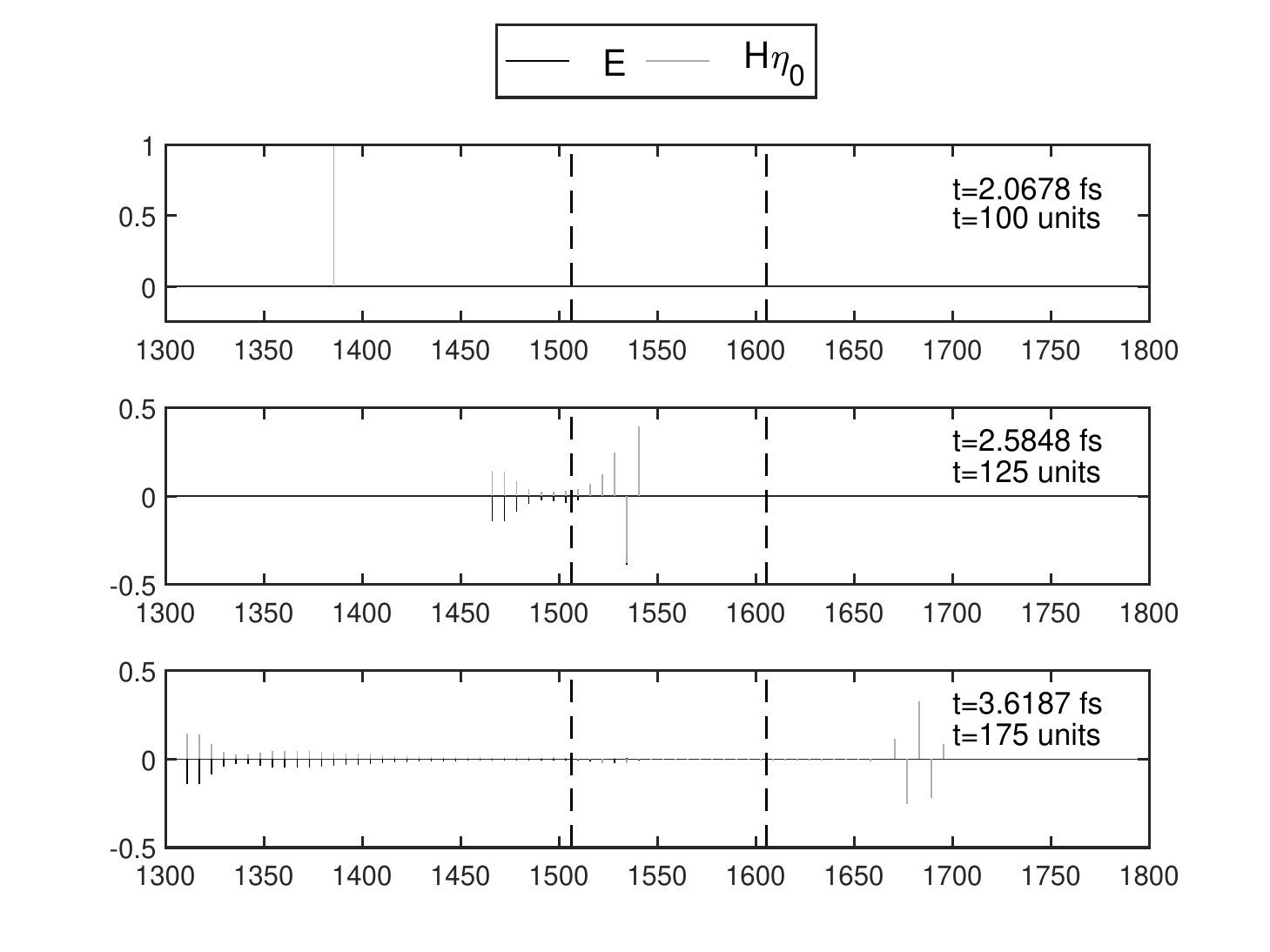}
    \caption{Transmission of electric and magnetic fields from a Dirac-Delta wave-function [source condition (3)] as it interacts with a silver slab of 100 \si{nm} thickness (boundaries represented by black-dashed lines). Axis units are in nanometers, \si{nm}.}
    \label{fig:5}
\end{figure}

\begin{figure}[!t]
    \centering
    \subfloat[]{\includegraphics[width=0.49\textwidth]{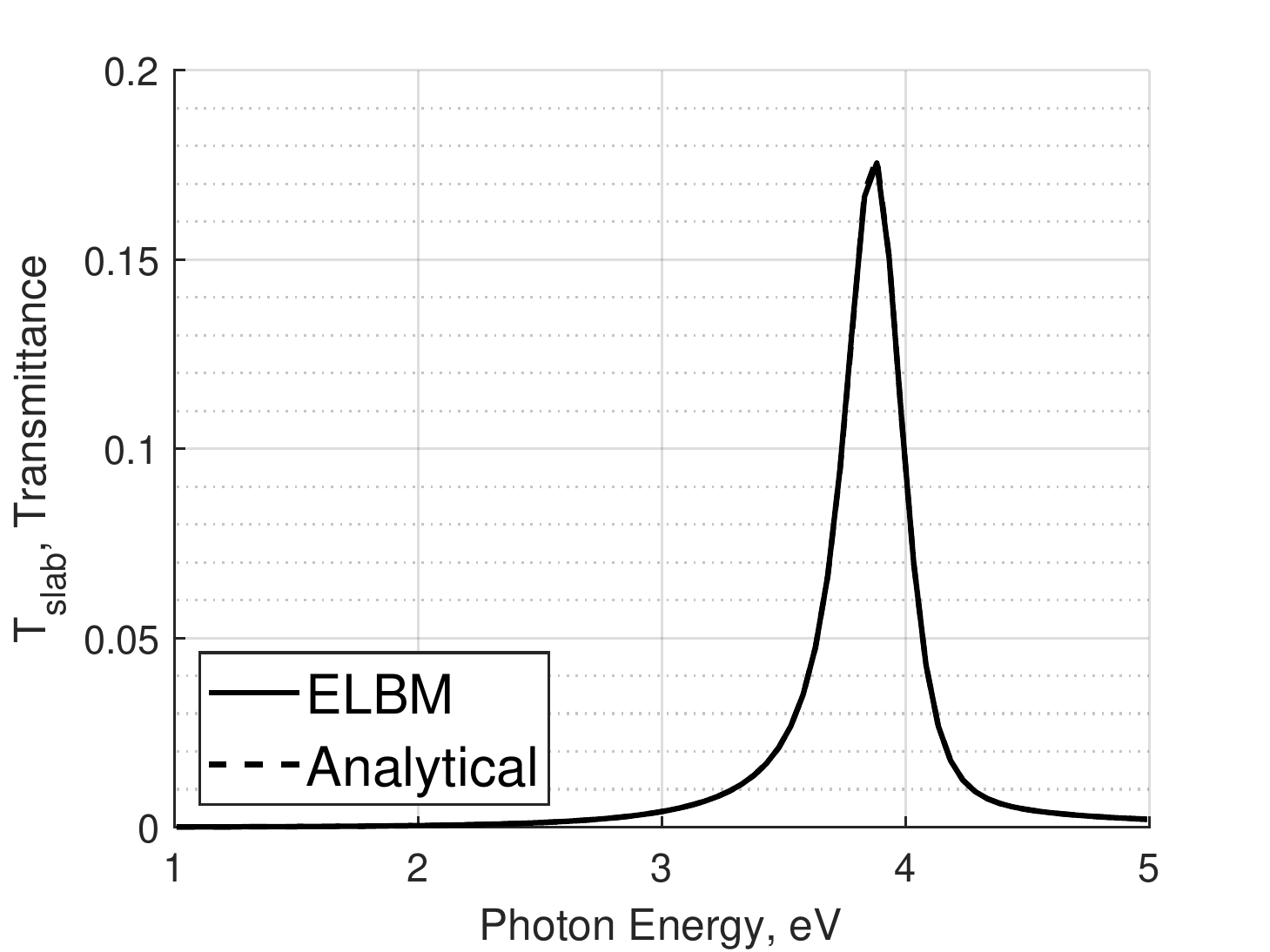}\label{fig:6a}}
    \subfloat[]{\includegraphics[width=0.49\textwidth]{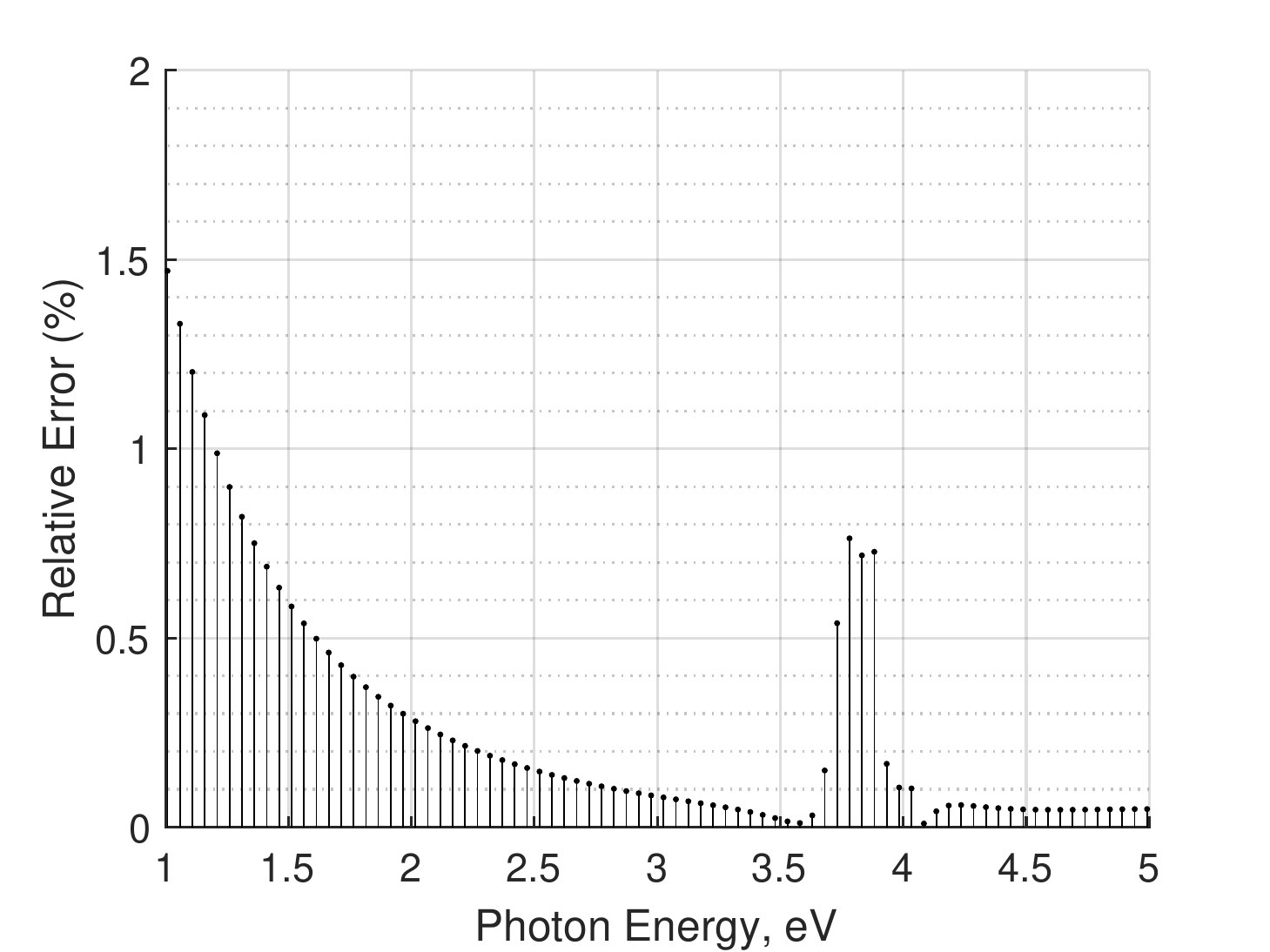}\label{fig:6b}}
    \caption{(a)  The 1D HV ELBM Dirac-delta wave-function's transmitted power spectral density compared with the analytical transmittance for a 100 \si{nm} thickness silver slab. The two are equivalent in this case. (b) The relative error in transmittance, with a mean of 0.28\% in the range 1-5 \si{eV} for $N_x=1000$ spatial lattice units, after $N_t=40000$ iterations for a computational time of 7.97 s. Simulation was performed on an 11\textsuperscript{th} Gen Intel\textsuperscript{\textregistered} Core\textsuperscript{TM} i7-11700K @ 3.60 GHz processor with 32 GB of RAM and 8 cores.}
    \label{fig:6}
\end{figure}

\begin{figure}[!t]
    \centering
    \subfloat[]{\includegraphics[width=0.49\textwidth]{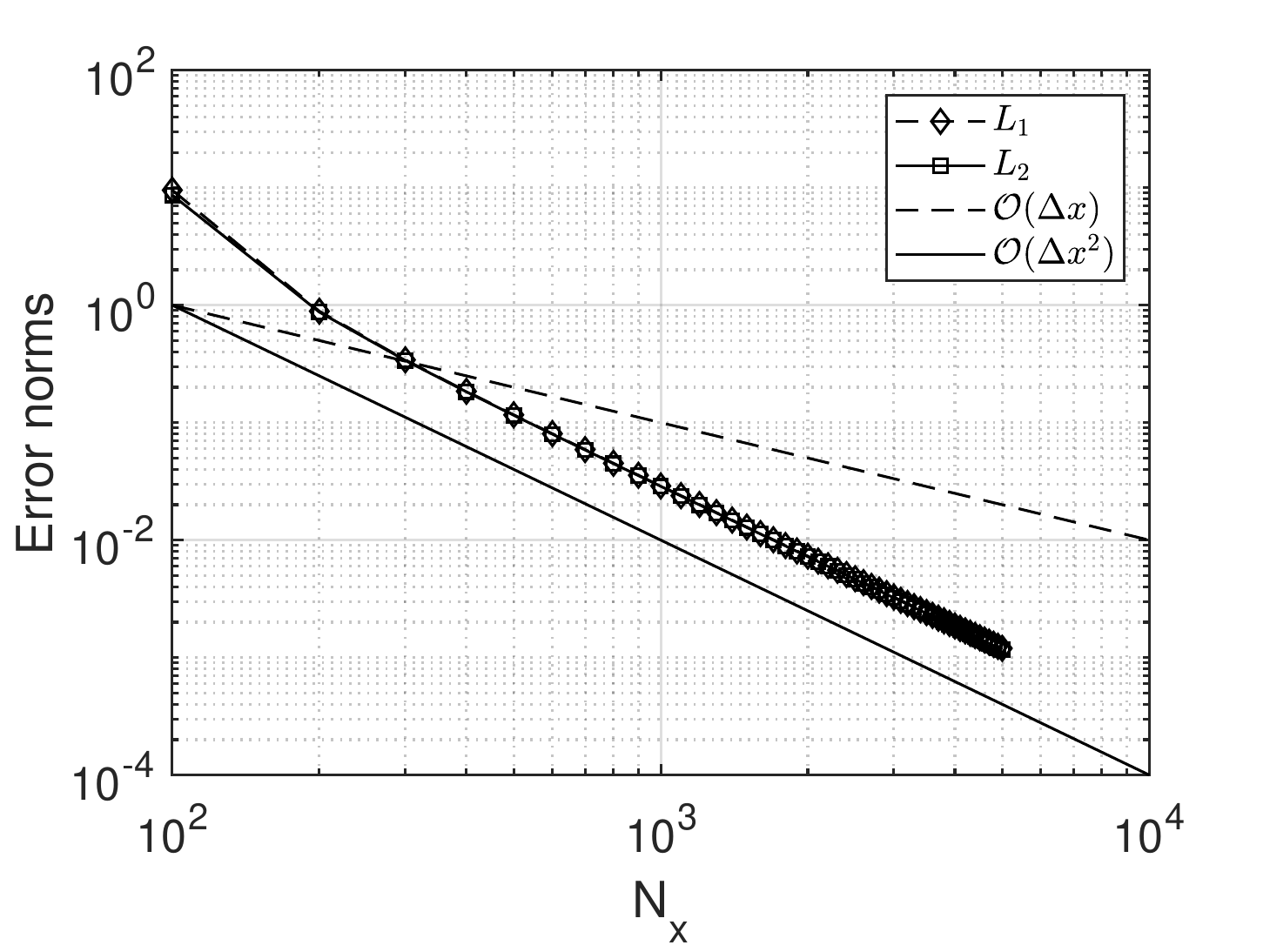}}
    \subfloat[]{\includegraphics[width=0.49\textwidth]{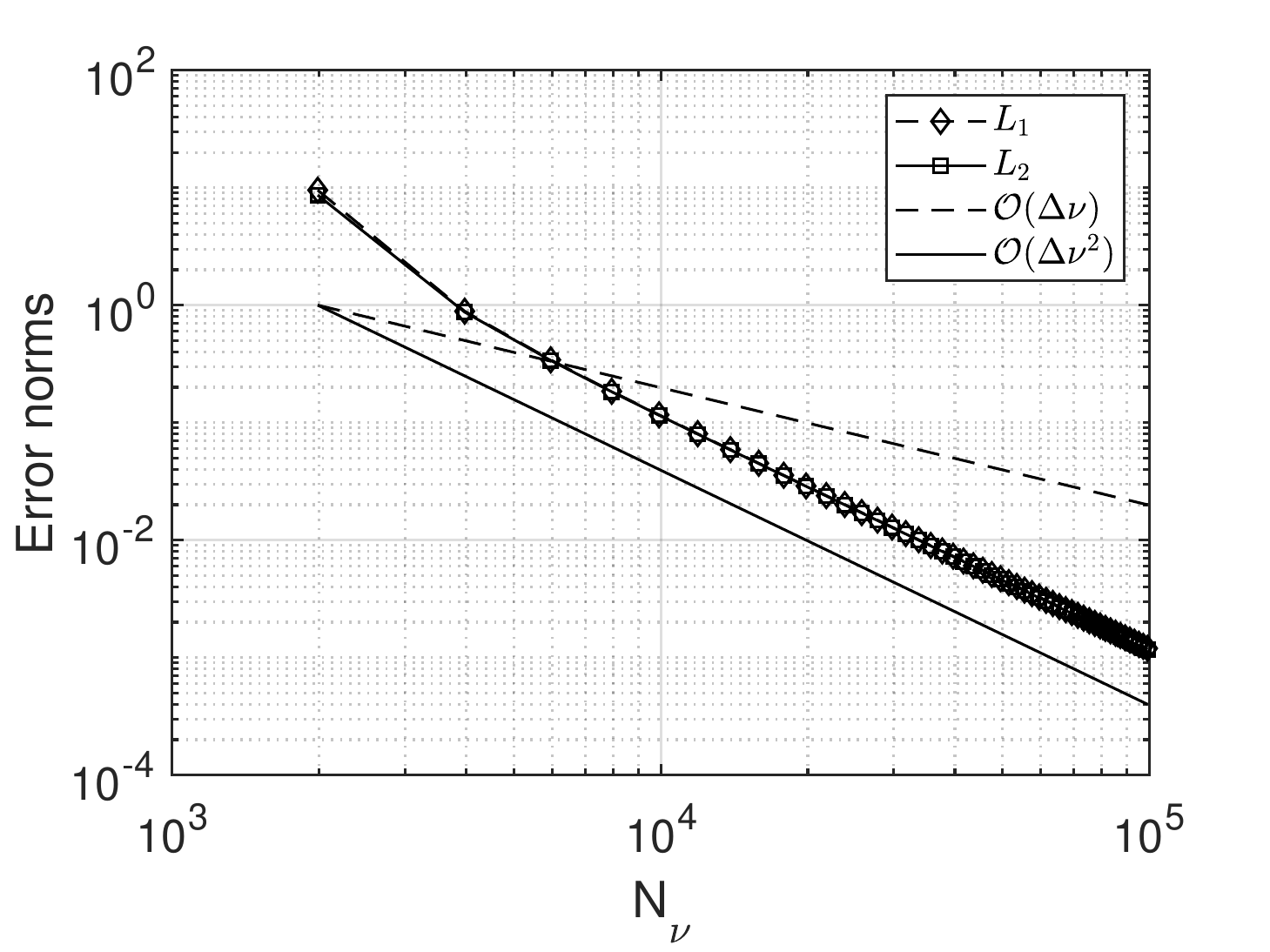}}
    \caption{ The relative error is plotted with a parallel slope to a second order of accuracy with respect to both (a) number of lattice units and (b) number of frequency bins. This CCPRP model requires $N_x>1100$ in order to obtain solutions within 1\% accuracy across the entire available spectrum (0-100 \si{eV} at $N_{x_1}=100$, 0-5000 \si{eV} at $N_{x_{50}}=5000$). This demonstrates that even a six pole CCPRP model is ill-suited for the HV ELBM's broad spectrum. The relative error is a more reliable measure for order-of-accuracy, given the large variation in the transmittance spectrum of silver. }
    \label{fig:7}
\end{figure}

Fig. \ref{fig:3} compares the relative errors, while Fig. \ref{fig:4} compares the computational time and memory, of the 1D HV ELBM and 1D FDTD method for source condition (2). While the two methods predict identical error norms from 1-5 \si{eV}, the 1D HV ELBM requires 60\% of the computational time for only 10\% more memory. Producing a propagating Dirac-delta wave-function in the 1D HV ELBM allows it to recover the broadband transmittance spectrum of a dispersive medium within a single simulation, which reduces the computational time by 2-3 orders of magnitude compared with the continuous-wave simulation.
Fig. \ref{fig:5} demonstrates the transmission of a Dirac-delta wave-function through the 100 \si{nm} thick silver slab in the 1D HV ELBM. The numerical stability of the simulation is maintained as the Dirac-delta wave-function transmits through the silver slab to represent its entire transmittance spectrum within the photon energy range of interest (1-5 \si{eV}). 
\newpage
\newpage
Fig. \ed{\ref{fig:6b}} shows a plot of the transmittance's relative error, which has a mean of 0.28\%  from 1-5 \si{eV}. Since the simulation was performed using $N_x=1000$ lattice units and $N_t=40000$ time units, the spectrum has a much greater bandwidth (0-1000 \si{eV}). While the 1D HV ELBM may represent frequencies up to the Nyquist limit, this requires developing a model with a greater number of CCPRPs representing the transmittance of silver over a broader spectrum. Nonetheless, a model fitted with six CCPRPs provides a reasonably accurate solution in the photon energy range of interest (1-5 \si{eV}), while remaining stable up to the Nyquist limit. Stability of the ADE allows calculating the relative error norms from the entire spectrum, for an accurate evaluation of the 1D HV ELBM order of accuracy.
The transmittance spectrum, evaluated from (\ref{eq:20}) for the case of the Dirac-delta wave-function transmitted through the 100 \si{nm} silver slab, is used to evaluate the 1D HV ELBM's order of accuracy across its entire bandwidth. The number of unit cells, $N_{x_k}$, is varied from $N_{x_1}=100$ ($\Delta x=6.2$ \si{nm}) to $N_{x_{50}}=5000$ ($\Delta x=0.124$ \si{nm}), while the number of time units, $N_{t_k}$, is varied from $N_{t_1}=4000$ ($\Delta t=20.7$ \si{as}), to $N_{t_{50}}=200000$ ($\Delta t=0.414$ \si{as}) to measure the first- and second-order relative error norms, $L_1$ and $L_2$, each with respect to both the number of lattice units, $N_x$, and the number of photon energy bins (frequency bins), $N_\nu$. The first- and second-order error norms have slopes that are parallel to the second-order relative error, as demonstrated in Fig. \ref{fig:7}. Therefore, the 1D HV ELBM in this study obtains a second-order accuracy and corroborates an earlier study in which the 1D HV ELBM's energy density achieved an absolute error of second-order accuracy for non-dispersive media \cite{Hauser2019}. Unlike absolute error \cite{Chen2013}, relative error is independent of the transmittance amplitude and is a preferable measure for the order-of-accuracy in this application. The ADE CCPRP model utilized by Han \emph{et al.} is applicable to photon energies ranging from 0.125-5 \si{eV} \cite{MinghuiHan2006}. Analyzing this model's transmittance over the 1D HV ELBM's complete spectrum (ranging from 0-100 \si{eV} at $N_{x_1}=100$ and $N_{\nu_1}=1986$, to 0-5000 \si{eV} at $N_{x_{50}}=5000$ and $N_{\nu_{50}}=99176$), we find its analytical transmittance amplitude increases significantly at photon energies exceeding 25 \si{eV}.  The relative error in Fig. \ref{fig:6} accounts for this variation in transmittance amplitude, and continues to decrease with a second-order slope. The error is significant for lower spatial discretizations, demonstrating that even a six-pole model is ill-suited for the broad spectrum represented by the 1D HV ELBM. Nonetheless, the relative error decreases to less than 1\% for $N_x>1100$, and demonstrates a second order-of-accuracy for the decreasing dimensional unit cell and time unit. Second order-of-accuracy is expected for the 1D HV ELBM model \cite{Hauser2019}, and the central difference equation used to incorporate the ADE \cite{MinghuiHan2006}. The second-order accurate 1D HV ELBM is an improvement from an earlier first-order accurate 1D ELBM for dispersive media \cite{Chen2013}.
\section{Conclusion}
The auxiliary-differential-equation with complex-conjugate pole-residue pairs can be added to the one-dimensional electrodynamic Hauser \& Verhey electrodynamic lattice-Boltzmann method (1D HV ELBM) to model the electromagnetic frequency dispersion of an evaporated silver slab. The implementation treats Debye and Lorentz media in a unified manner while recovering electric and magnetic fields similar to the one-dimensional finite-difference-time-domain (1D FDTD) method. Compared with the 1D FDTD method for the same lattice unit and accuracy, the 1D HV ELBM requires 60\% the computational time and only 10\% greater memory. However, this is not the main advantage of the 1D HV ELBM. Since the 1D HV ELBM predicts the stable transmission of a Dirac-delta wave-function through silver, it can describe its transmittance spectrum within a single simulation. Therefore, given a sufficiently stable complex-conjugate pole-pair model for a dispersive medium, we find the 1D HV ELBM is a powerful alternative to the 1D FDTD method for modeling the broadband frequency dispersion of electromagnetic waves.

\bibliographystyle{IEEEtran}
\bibliography{ELBM_dispersive_v2}  






\end{document}